# A Body Area Network through Wireless Technology

Ramesh GP[1], Aravind CV[2], Rajparthiban R[3], N.Soysa[4]

[1]St.Peter's University, Chennai, India
[2]Computer Intelligence Applied Research Group, School of Engineering, Taylor's University, Malaysia
[3]The University of Nottingham Malaysia Campus, Semeniy, Malaysia
[4]UCSI University Cheras Malaysia

Abstract: A physiological signal monitoring system and alerting system using wireless technology is presented. The two types of physiological signal monitoring are captured from the body through leads and using the radio-frequency transmitting and receiving module the data are interfaced to computer systems. Furthering using a developed user interface module the captured signals are analyzed for checking abnormality. Any significant recordings are transmitted to the physicians hand phone by using external serial SMS modem. ECG signal de-noising is conducted by using low-pass and high-pass filters. EEG signals de-noising is conducted by using band-pass filters set. A comparative evaluation of the module with the manual recording shows encouraging results. The ECG and EEG pattern are presented in this paper.

I. INTRODUCTION

Wearable health monitoring systems integrated into a telemedicine system are novel information technology that will be able to support early detection of abnormal conditions and prevention of its serious consequences [1,2]. A continuous monitoring diagnostic procedure of a chronic condition or during supervised recovery from an acute event or surgical procedure is needed these days. Wireless Body Area Network, (WBAN) consists of a set of mobile and compact intercommunicating sensors, either wearable or implanted into the human body, which monitor vital body parameters and movements [3-5]. This paper aims to present the design and implement of a portable device that can capture ECG (Electrocardiogram) and EEG (Electroencephalography) signals from the human body and send those signals into Personal Computer (PC) for continuous monitoring and analysis. In the event of any abnormal changes of in the captured signal, the PC sends a message to doctor's hand phone. ECG and EEG signals are captured from the electrodes and it is send to the portable device that fixed on body. Here these signals will be amplified before sending it to RF transmitting device. Transmitted signal is collected from the receiving module and the signals through a low pass filter before recording in the database for analyzing.

Inside the developed program of the software ECG signals are de-noised using low-pas filter (FIR) and low-pass filter (FIR). Using de-noised ECG signals, heart beat and amplitude of the R intervals are analyzed. EEG signal are de-noised by using band pass-filters (FIR). EEG signals are then categorized according to the frequency ranges. This system was mainly designed to analyze alpha-wave (α-wave). Leads and the potentials recorded at the various point in the test object is as shown in Fig.1 The physiological signals are of low amplitude signals; therefore it is very important to amplify these signals before being transmitted. Instrumentation amplifier can be used to amplify these mV range physiological signals. Fig.2 shows the block diagram representation of the set-up.

II. TEST SET-UP

*2.1 Electrocardiogram Testing*

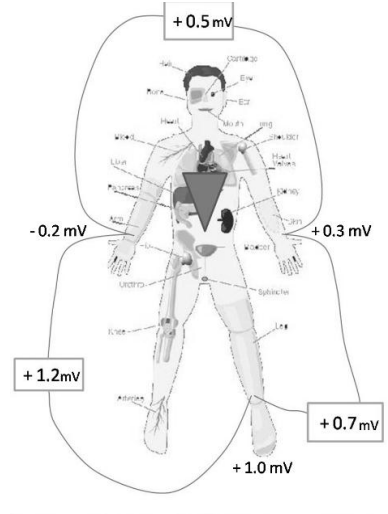

Fig.1. Position of the Lead and the maximum voltage level

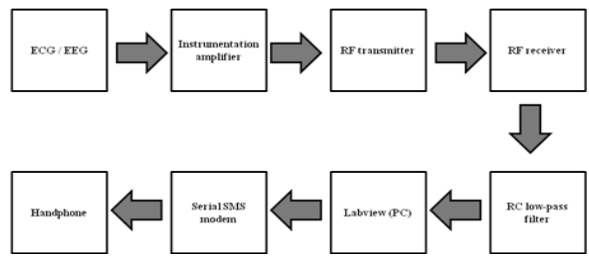

Fig. 2. Block diagram of the working system

Lead one signal was captured by placing electrodes in left hand, right hand and left leg signal was connected to the common ground. Figure 3(a). shows the signal that was collected by placing the left leg signal in common ground and Figure 3(b) represents the signal that was captured without placing left leg signal to the common ground. Lead two signals were captured by placing electrodes on the right arm and left leg and left arm was connected to the common ground. Figure 3(c). shows signals that were captured with common ground and Figure 3(d) shows signals that were captured without common ground. Lead three signals were captured by placing electrodes on the left leg, left arm and the right arm was connected to the common ground. Figure 3(e). shows signals that were captured with common ground and Figure 3(f) shows signals that were captured without common ground.

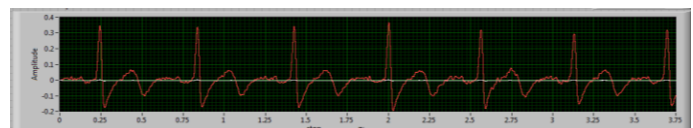

Fig. 3(a). Lead 1 voltage level measurement

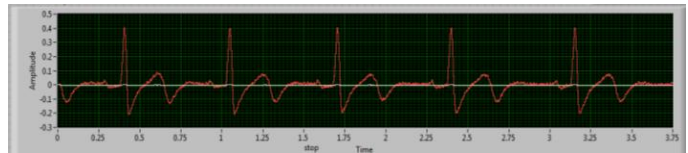
Fig. 3(b). Lead 1 voltage level measurement

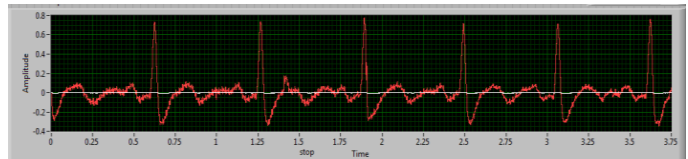
Fig.3 (c). Lead 2 voltage level measurements

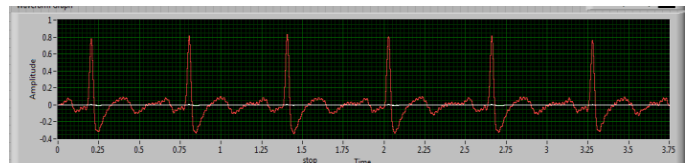
Fig.3 (d). Lead 2 voltage level measurements

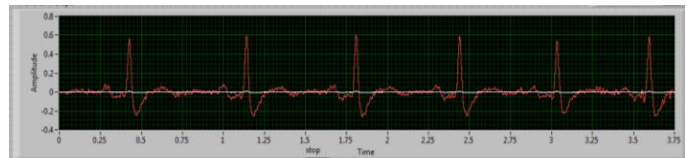
Fig. 3(e). Lead 3 voltage level measurement

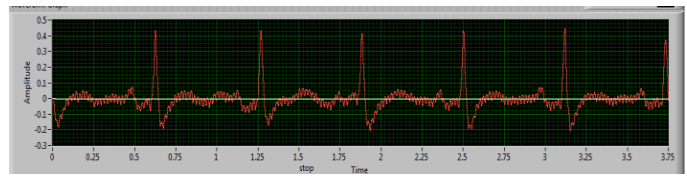
Fig. 3(f). Lead 3 voltage level measurement

## *2.2 Electroencephalography Testing*

EEG Signals are very smaller compare to the ECG signals. It is important to amplify these uV level EEG signal before insert into PC for further processing and analyze. Before amplify above signals into higher voltage it is important to reduced the noisy signal that contain with the EEG signal so noisy EEG signal collected by electrodes send through RC band-pass filter to reduced the noise level. Signal De noising will increase the SNR. Figure 4 shows the block diagram representation of the set-up. To capture EEG signals electrodes were placed in left and right side of the forehead. These two electrode signal send through the RC filter and send into the amplifier for amplify. Third electrode was fixed to left side of the neck and it was connected with common ground of the circuit. Figure 5 shows the de-noised ECG signal.

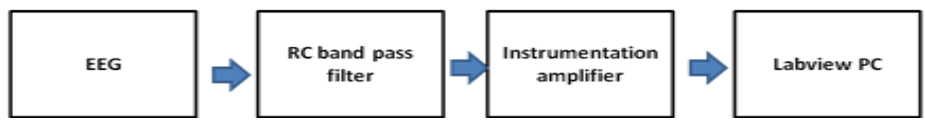
Fig.4(a) Block diagram of the EEG signal

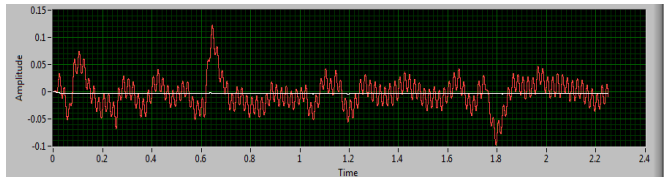

Fig.4(b) De-noised EEG Signal

### III. SOFTWARE DEVELOPMENT

A GUI as shown in Figure 5(a) is developed for the human machine interface including provisions for the ECG/EEG and alarm setup for abnormality recordings. The signal fed through the microphone port of the PC was interfaced with Labview by using sound acquiring block set. Using FIR digital high-pass filter, low frequency noisy signals was eliminated from the ECG signal. FIR High-pass filter is used to eliminate the entire high frequency noisy signal that left in ECG signals. To calculate the heart beat rate, tone measurement blackest was used to find out frequency of the ECG signal and then frequency was multiplied by frequency. Noisy EEG signal fed into PC was de-noised by FIR band-pass filter. To measure frequency of the EEG signal tone measurement block set was used. Figure 5(b) shows the human machine interface through instrumentation tool.

$$HB = f_{ECG} * f \qquad (1)$$

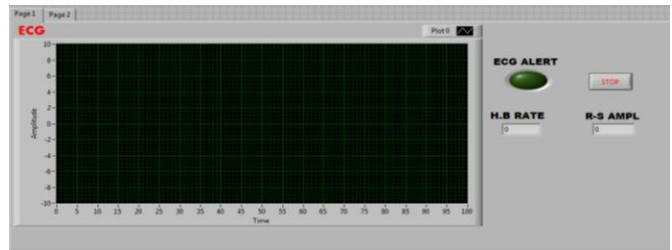

Fig. 5(a) Human Machine Interface for ECG

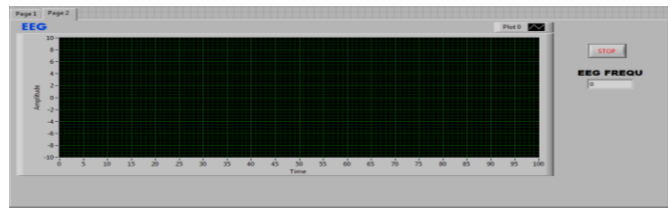

Fig. 5(b) Human Machine Interface for EEG

### IV. RESULTS

Low-pass filter output results shows that most of the unwanted noisy signals have been roved from the ECG signals. According to results, ECG signal P, Q, R, S and T point can be clearly observed. The results been achieved shows some abnormality in U peak of the ECG signal and it occurs regularly throughout the waveform. It's proved this abnormality occurs not due to the noisy signals. This abnormality of the signals occurs due to, signal was captured by placing only three electrodes on the body to get the perfect ECG signal it is important place nine electrodes on the body. Figure 6 ECG results were collected when the tester was in sitting position. Electrodes were placed in right arm left arm and left leg. ECG signals that been captured and amplified by instrumentation amplifier transmitted by using walkie-talkie module. Figure 6(d) represents transmitted and de-noised ECG signal. According to Figure 6(d) results, observed some major changes occur in ECG signal. Above ECG signal Q, R and S peaks can be clearly seen, but the P, T and U peaks can't be seen. According to the results achieved, it's proven that wireless and wire ECG signal has the same frequency.

When conceder about the phase of the signal we can observe that some phase distortions has been occurred. Phase distortions occurs due to filter type that been used in walkie-talkie module to de-noise sound signals.

TABLE 1 ECG Recordings

|  | Amplitude Difference (mV) | |
| --- | --- | --- |
|  | ECG Machine | Using the Module |
| **Lead 1** | | |
| P | 0.2 | 0.2 |
| R->S | 0.55 | 0.53 |
| T | 0.3 | 0.25 |
| **Lead 2** | | |
| P | 0.3 | 0.38 |
| R->S | 1.4 | 1.1 |
| T | 0.5 | 0.45 |
| **Lead 3** | | |
| P | 0.2 | 0.25 |
| R->S | 0.8 | 0.8 |
| T | 0.3 | 0.28 |

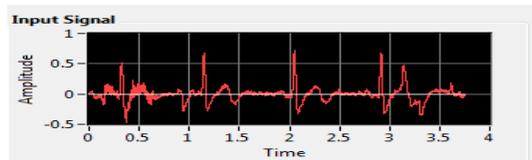
Fig.6(a) Input signal that enter into the High-pass filter

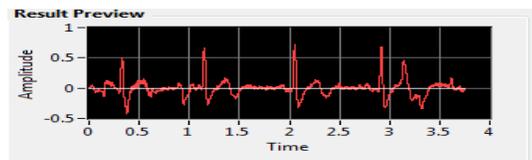
Fig.6(b) Output signal that from the high-pass filter

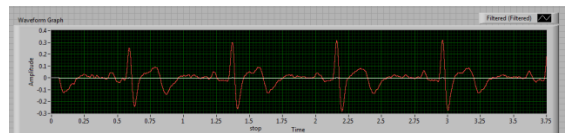
Fig.6(c) Output signal from the low-pass filter

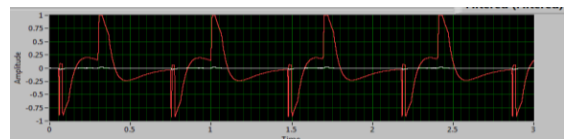
Fig. 6(d) Transmitted ECG signal

Figure 7 shows the ECG recordings, the EEG results was collected when the person was in relaxing passion. Electrodes were placed on right side and left sides of the forehead and the third electrode was placed in left side of the neck. Band-pass filter output results was shown that the most of the unwanted noisy signal been removed from the EEG signal.

According to the result been achieved shows some sudden low and high peaks in the signal these abnormalities occurs due to EMG signals interference with EEG signals and also this signal was captured by placing only three electrodes and to capture EEG signal ECG electrodes was used it can cause some errors in captured signal (Note: in general, to capture EEG and ECG, Ag/Agcl type electrodes used). Tone measurement block set was used to analyze frequency range of the EEG signal. The EEG frequency was recorded as range of 7-13Hz. Frequency range was proven that captured EEG signal is alpha wave.

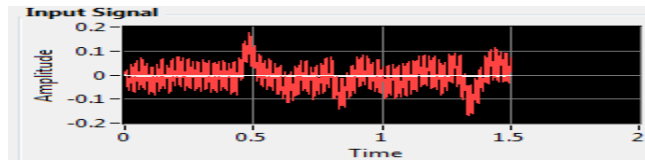

Fig.8(a) input signal that enter into the band-pass filter

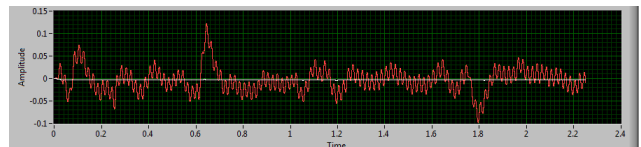

Fig.8(b) Output signal that from the band-pass filter

V. CONCLUSIONS

Main aim of the paper is to design a system that can monitor human body signals and Diogenes health condition of sic person. To achieve this, it is important to have good idea about the ECG and EEG signals characteristics and the behavior of the signals according to changes in health situation of the person. It is also important to know the characteristics of the filters, RF transmitter modules, Labview software, instrumentation amplifier and SMS serial modem. According to the results obtained, this system can be helpful for doctors to closely monitor their patients throughout the day.